\documentclass[aps,prl,showpacs,twocolumn]{revtex4}
\usepackage{amsmath, amssymb}

\begin{document}

\title{Quaternions, octonions and Bell-type inequalities}

\author{E. Shchukin}
\email{evgeny.shchukin@gmail.com}
\author{W. Vogel}
\email{werner.vogel@uni-rostock.de}
\affiliation{Arbeitsgruppe Quantenoptik, Institut f\"{u}r  Physik,
Universit\"at Rostock, D-18051 Rostock, Germany}

\begin{abstract}
Multipartite Bell-type inequalities are derived for general systems. They involve up to eight observables with arbitrary spectra on each site. These inequalities are closely related to the algebras of quaternions and octonions. 
\end{abstract}


\maketitle

In their famous paper Einstein, Podolsky and Rosen (EPR) suggested a Gedankenexperiment~\cite{pr-47-777}. As they believed, it would prove the incompleteness of quantum mechanics. An interesting analysis of this
problem was given by Bohr \cite{pr-48-696}, who did not agree with EPR. Almost thirty years later, in $1964$, in his remarkable paper Bell proposed a quantitative test which should resolve the problem of completness of quantum mechanics~\cite{ph-1-195}. He showed that
the assumptions of the EPR arguments lead to some restrictions on multipartite correlations, which are now referred to as Bell inequalities. If Bohr's arguments are correct, then these inequalities can be violated. In the $1980$'s the first experimental violations of these inequalities were demonstrated~\cite{prl-45-617, n-398-189}, and thus the arguments of Bohr were verified. 

There exist many different Bell-type inequalities. Most of them deal with dichotomic observables or with generalizations to observables with more general discrete spectra, for example see \cite{prl-88-040404}. On the other hand, in the original EPR paper the situation of observables with a continuous spectrum was considered. For the case of observables with a general spectrum, to this end the theory of Bell-type inequalities is much less developed. 

The first multipartite Bell-type inequality, valid for arbitrary observables, has been obtained very recently~\cite{prl-99-210405}. 
In the simplest, bipartite, case with two observables $\hat{A}_j$, $\hat{B}_j$ on each site ($j=1,2$) it reads as
\begin{equation}\label{eq:b2}
\begin{split}
    &|\langle(\hat{A}_1+i\hat{B}_1)(\hat{A}_2+i\hat{B}_2)\rangle|^2 =
    \langle \hat{A}_1\hat{A}_2 - \hat{B}_1\hat{B}_2 \rangle^2 \\
    &+\langle \hat{B}_1\hat{A}_2 + \hat{A}_1\hat{B}_2 \rangle^2 \leqslant 
    \langle (\hat{A}^2_1 + \hat{B}^2_1)(\hat{A}^2_2 + \hat{B}^2_2) \rangle.
\end{split}
\end{equation}
The original proof is based on ignoring local commutators in the quantum mechanical analogue of a classical inequality. In some sense such a procedure is ambiguous, since only those commutators appearing explicitly are ignored. Furthermore, no direct proof was given that  the resulting inequality is fulfilled for any separable quantum state -- only in this case it should be called Bell-type. 

In this letter we give a strict and unambiguous proof of general Bell-type inequalities, including their relation to the separability
problem. It allows us to consider up to eight arbitrary observables at each site, for a complex quantum system. The multipartite extension of the inequalities is based on the algebra of quarternions and octonions. The  recently proposed inequalities~\cite{prl-99-210405} will appear as special cases of our approach. 

Let us start to prove that each bipartite separable state satisfies the inequality~\eqref{eq:b2}. This approach also allows us to obtain Bell-type inequalities with four and eight observables on each site. First, we prove a more general statement: if $\hat{F}_1, \ldots, \hat{F}_k$ and $\hat{G}$ are Hermitian operators such that some (in general, multipartite) states $\hat{\varrho}_j$, $j = 0, 1, \ldots$,  satisfy the inequality
\begin{equation}\label{eq:FFF}
    \langle\hat{F}_1\rangle^2 + \ldots + \langle\hat{F}_k\rangle^2 \leqslant \langle\hat{G}\rangle,
\end{equation}
then arbitrary mixtures (i.e. convex combinations) of these states also satisfy this inequality. The proof is based on the Cauchy-Schwarz inequality
\begin{equation}\label{eq:cse}
    \Bigl|\sum^{+\infty}_{j=0} x^*_j y_j\Bigr|^2 \leqslant 
    \Bigl(\sum^{+\infty}_{j=0} |x_j|^2\Bigr) \Bigl(\sum^{+\infty}_{j=0} |y_j|^2\Bigr),
\end{equation}
which is valid for all complex numbers $x_j$ and $y_j$. The case of $y_j \equiv p_j$ being a probability distribution, i.e. $p_j \geqslant 0$ and $\sum^{+\infty}_{j=0} p_j = 1$, is of special interest. Then one gets from the inequality \eqref{eq:cse} 
\begin{equation}\label{eq:px}
    \Bigl|\sum^{+\infty}_{j=0} p_j x_j\Bigr|^2 \leqslant \sum^{+\infty}_{j=0} p_j |x_j|^2.
\end{equation}
Note that the sum on the left hand side of this inequality is the mathematical expectation of the random variable $X$ that attains the value $x_j$ with probability $p_j$, and the right hand side is the mathematical expectation of the square of this random variable. The inequality \eqref{eq:px} then states that variance of $X$, $\sigma_X$, is nonnegative.

Let us now take a convex combination $\hat{\varrho} = \sum^{+\infty}_{j=0} p_j \hat{\varrho}_j$ and estimate the left hand side of the inequality \eqref{eq:FFF}. According to the inequality \eqref{eq:px} we have
\begin{equation}
    \langle\hat{F}_l\rangle^2 = \Bigl(\sum^{+\infty}_{j=0} p_j \langle\hat{F}_l\rangle_j\Bigr)^2 \leqslant
    \sum^{+\infty}_{j=0} p_j \langle\hat{F}_l\rangle^2_j,
\end{equation}
where $\langle\hat{F}_l\rangle_j$ is the average value on the state $\hat{\varrho}_j$. Now we can estimate the left hand side of \eqref{eq:FFF} as follows:
\begin{equation}
    \sum^k_{l=1} \langle\hat{F}_l\rangle^2 \leqslant \sum^{+\infty}_{j=0} p_j \sum^k_{l=1} 
    \langle\hat{F}_l\rangle^2_j \leqslant \sum^{+\infty}_{j=0} p_j \langle\hat{G}\rangle_j = 
    \langle\hat{G}\rangle.
\end{equation}
So, we have obtained the desired result. 

Now we can easily prove the inequality \eqref{eq:b2} for separable states. Here we have $k=2$ and $\hat{F}_1 = \hat{A}_1\hat{A}_2 - \hat{B}_1\hat{B}_2$, $\hat{F}_2 = \hat{B}_1\hat{A}_2 - \hat{A}_1\hat{B}_2$ and $\hat{G} = (\hat{A}^2_1 + \hat{B}^2_1)(\hat{A}^2_2 + \hat{B}^2_2)$. The have to show that the inequality \eqref{eq:b2} is valid for all factorizable
states. Then we get
\begin{equation}\label{eq:eq}
\begin{split}
    \langle \hat{A}_1\hat{A}_2 &- \hat{B}_1\hat{B}_2 \rangle^2 + 
    \langle \hat{B}_1\hat{A}_2 + \hat{A}_1\hat{B}_2 \rangle^2 = \\ 
    (\langle\hat{A}_1\rangle^2 &+ \langle\hat{B}_1\rangle^2)
    (\langle\hat{A}_2\rangle^2 + \langle\hat{B}_2\rangle^2) \leqslant \\
    &\langle (\hat{A}^2_1 + \hat{B}^2_1)(\hat{A}^2_2 + \hat{B}^2_2) \rangle.
\end{split}
\end{equation}
The equality in this chain is valid since only squares remain in the sum and mixed terms cancel each other. The last step, the inequality, is valid since it just expresses the fact the variance of an observable is nonnegative. We conclude that each convex combination of factorizable state, i.e. each separable state satisfies the inequality \eqref{eq:b2}.

We see that the key point in our proof of the inequality \eqref{eq:b2} is the estimation \eqref{eq:eq}, which can be divided into two steps. The first step, the equality, can be expressed as the following square identity:
\begin{equation}\label{eq:sq2}
    (a_1 a_2 - b_1 b_2)^2 + (a_1 b_2 + b_1 a_2)^2 = (a^2_1 + b^2_1)(a^2_2 + b^2_2).
\end{equation}
The second step is valid due to the nonnegativity of the variance of observables. Having an identity of the form 
\begin{equation}\label{eq:sq}
    (a^2_1 + b^2_1 + \ldots)(a^2_2 + b^2_2 + \ldots) = x^2 + y^2 + \ldots,
\end{equation}
where all the sums contain the same number $n$ of terms and $x$, $y$, \ldots are bilinear functions of $a_l$, $b_l$, \ldots, $l = 1, 2$, we can immediately write a Bell-type inequality
\begin{equation}
    \langle \hat{X} \rangle^2 + \langle \hat{Y} \rangle^2 + \ldots \leqslant 
    \langle (\hat{A}^2_1 + \hat{B}^2_1 + \ldots)(\hat{A}^2_2 + \hat{B}^2_2 + \ldots) \rangle,
\end{equation}
where $\hat{X}$, $\hat{Y}$, \ldots are the Hermitian operators obtained by replacing $a_l$, $b_l$, \ldots by arbitrary Hermitian operators $\hat{A}_l$, $\hat{B}_l$, \ldots, $l = 1, 2$ in the bilinear forms $x$, $y$, \ldots respectively. Let us again formulate the reasons why this inequality is valid for all separable states. Firstly, it is of the form \eqref{eq:FFF}, i.e. if it is valid for some states, it is also valid for their mixtures. Secondly, it is valid for all factorizable states due to the identity \eqref{eq:sq} and nonnegativity of the variances of observables. It follows that it is valid for all mixtures of any factorizable states, i.e. for all separable states.

What square identities exist? The case of $n=2$ was considered above. There are also square identities for $n=4$ and $n=8$. They read as: Euler four square identity
\begin{equation}\label{eq:sq4}
\begin{split}
    &(a_1 a_2 - b_1 b_2 - c_1 c_2 - d_1 d_2)^2 + \\
    &(b_1 a_2 + a_1 b_2 - d_1 c_2 + c_1 d_2)^2 + \\
    &(c_1 a_2 + d_1 b_2 + a_1 c_2 - b_1 d_2)^2 + \\
    &(d_1 a_2 - c_1 b_2 + b_1 c_2 + a_1 d_2)^2 = \\
    (a^2_1 &+ b^2_1 + c^2_1 + d^2_1)(a^2_2 + b^2_2 + c^2_2 + d^2_2),
\end{split}
\end{equation}
and Degen eight-square identity
\begin{widetext}\small
\begin{equation}\label{eq:sq8}
\begin{split}
    &(a_1 a_2 - b_1 b_2 - c_1 c_2 - d_1 d_2 - e_1 e_2 - f_1 f_2 - g_1 g_2 - h_1 h_2)^2 + 
    (b_1 a_2 + a_1 b_2 + d_1 c_2 - c_1 d_2 + f_1 e_2 - e_1 f_2 - h_1 g_2 + g_1 h_2)^2 + \\
    &(c_1 a_2 - d_1 b_2 + a_1 c_2 + b_1 d_2 + g_1 e_2 + h_1 f_2 - e_1 g_2 - f_1 h_2)^2 + 
    (d_1 a_2 + c_1 b_2 - b_1 c_2 + a_1 d_2 + h_1 e_2 - g_1 f_2 + f_1 g_2 - e_1 h_2)^2 + \\
    &(e_1 a_2 - f_1 b_2 - g_1 c_2 - h_1 d_2 + a_1 e_2 + b_1 f_2 + c_1 g_2 + d_1 h_2)^2 + 
    (f_1 a_2 + e_1 b_2 - h_1 c_2 + g_1 d_2 - b_1 e_2 + a_1 f_2 - d_1 g_2 + c_1 h_2)^2 + \\
    &(g_1 a_2 + h_1 b_2 + e_1 c_2 - f_1 d_2 - c_1 e_2 + d_1 f_2 + a_1 g_2 - b_1 h_2)^2 + 
    (h_1 a_2 - g_1 b_2 + f_1 c_2 + e_1 d_2 - d_1 e_2 - c_1 f_2 + b_1 g_2 + a_1 h_2)^2 = \\
    &(a^2_1+b^2_1+c^2_1+d^2_1+e^2_1+f^2_1+g^2_1+h^2_1)(a^2_2+b^2_2+c^2_2+d^2_2+e^2_2+f^2_2+g^2_2+h^2_2).
\end{split}
\end{equation}
\end{widetext}
The famous Hurwitz theorem states that there are no other identities of such a form \cite{quaternions}.

The inequality corresponding to the identity \eqref{eq:sq4} is
\begin{equation}\label{eq:b4}
\begin{split}
    &\langle\hat{A}_1 \hat{A}_2 - \hat{B}_1 \hat{B}_2 - \hat{C}_1 \hat{C}_2 - \hat{D}_1 \hat{D}_2\rangle^2 + \\
    &\langle\hat{B}_1 \hat{A}_2 + \hat{A}_1 \hat{B}_2 - \hat{D}_1 \hat{C}_2 + \hat{C}_1 \hat{D}_2 \rangle^2 + \\
    &\langle\hat{C}_1 \hat{A}_2 + \hat{D}_1 \hat{B}_2 + \hat{A}_1 \hat{C}_2 - \hat{B}_1 \hat{D}_2\rangle^2 + \\
    &\langle\hat{D}_1 \hat{A}_2 - \hat{C}_1 \hat{B}_2 + \hat{B}_1 \hat{C}_2 + \hat{A}_1 \hat{D}_2\rangle^2 \leqslant \\
    \langle(\hat{A}^2_1 &+ \hat{B}^2_1 + \hat{C}^2_1 + \hat{D}^2_1)
    (\hat{A}^2_2 + \hat{B}^2_2 + \hat{C}^2_2 + \hat{D}^2_2)\rangle
\end{split}
\end{equation}
This is a bipartite Bell-type inequality with four observables on each site. To extend it to the general multipartite case it is natural to use the algebra of quaternions. Remember that this algebra has dimension $4$ over the reals, so each quaternion $q$ can be written as $q = x + i y + ju + k v$ in an unique way, where $x$, $y$, $u$ and $v$ are reals. The multiplication rules for the imaginary units $i$, $j$ and $k$ are $i^2 = j^2 = k^2 = -1$, $ij = -ji = k$, $jk = -kj = i$, $ki = -ik = j$. The conjugation $q^*$ of the quaternion $q$ is defined as $q^* = x - i y - ju - k v$. The norm of $q$ is defined in the standard way as $|q| = \sqrt{q^* q}$. The identity \eqref{eq:sq4} simply express the fact that the norm is multiplicative: $|q^{\prime}q^{\prime\prime}| = |q^{\prime}| |q^{\prime\prime}|$ for arbitrary quaternions $q^{\prime}$ and $q^{\prime\prime}$. The norm also satisfies the triangle inequality $|q_1 + q_2| \leqslant |q_1| + |q_2|$. The inequality \eqref{eq:b2} is formulated using the operators of the form $\hat{f} = \hat{A}+i\hat{B}$, which is a general form of a non-Hermitian operator. We extend the class of operators acting on the state space of the system to quaternionic operators of the form $\hat{q} = \hat{A} + i \hat{B} + j \hat{C} + k \hat{D}$, where $\hat{A}$, $\hat{B}$, $\hat{C}$ and $\hat{D}$ are ordinary Hermitian operators. Since the algebra of quaternions is noncommutative, care must be taken when defining the product of quaternions with operators. We define this product such that, if $\hat{q}_m = \hat{A}_m + i \hat{B}_m + j \hat{C}_m + k \hat{D}_m$ are quaternionic operators acting on different degrees of freedom $m = 1, \ldots, n$, then
\begin{equation}\label{eq:qq}
    \langle\hat{q}_1 \ldots \hat{q}_n\rangle = \langle\hat{q}_1\rangle \ldots \langle\hat{q}_n\rangle
\end{equation}
for each completely factorizable state.	Let us calculate the average value of the product $j \hat{f}$, where the operator $\hat{f}$ has been defined above
\begin{equation}
\begin{split}
    \langle j\hat{f}\rangle &= j\langle \hat{f}\rangle = j(\langle\hat{A}\rangle + i\langle\hat{B}\rangle) = 
    j\langle\hat{A}\rangle - ij\langle\hat{B}\rangle \\
    &= \langle\hat{A}-i\hat{B}\rangle j = \langle\hat{f}^\dagger j\rangle.
\end{split}
\end{equation}
Here we used the fact that the numbers $\langle\hat{A}\rangle$ and $\langle\hat{B}\rangle$ are real and due to this they commute with $j$. The same is valid with respect to the other imaginary unit $k$. We see that the natural way to define the product of the quaternion $q$ with the operator $\hat{f}$ is $q \hat{f} = \hat{f}(x + i y) + \hat{f}^\dagger (j u + k v)$. In particular, if $\hat{F}$ is a Hermitian operator, then we have $q\hat{F} = \hat{F}q$, so the quaternionic operators $\hat{q}$ defined above behave as if the operators $\hat{A}$, $\hat{B}$, $\hat{C}$ and $\hat{D}$ were ordinary  real numbers. This guarantees that the equality \eqref{eq:qq} is fulfilled for all factorizable states.

The same idea we used to prove the inequality \eqref{eq:FFF} allows us to prove the following statement: if $\hat{q}_m$ are quaternionic operators and $\hat{F}_m$ are Hermitian operators, acting on different degrees of freedom, such that $|\langle\hat{q}_m\rangle|^2 \leqslant \langle\hat{F}_m\rangle$, $m = 1, \ldots, n$, then each completely separable state satisfies the inequality
\begin{equation}\label{eq:qF}
    |\langle\hat{q}_1 \ldots \hat{q}_m\rangle|^2 \leqslant \langle\hat{F}_1 \ldots \hat{F}_n\rangle.
\end{equation}
Here we need the factorization property \eqref{eq:qq}, multiplicativity of the norm and the triangle inequality. Since we can estimate $|\langle\hat{q}_m\rangle|^2$ as
\begin{equation}
\begin{split}
    |\langle\hat{q}_m\rangle|^2 &= \langle\hat{A}_m\rangle^2+\langle\hat{B}_m\rangle^2+\langle\hat{C}_m\rangle^2+\langle\hat{D}_m\rangle^2 \\ 
    &\leqslant \langle\hat{A}^2_m+\hat{B}^2_m+\hat{C}^2_m+\hat{D}^2_m\rangle,
\end{split}
\end{equation}
we can take $\hat{F}_m = \hat{A}^2_m+\hat{B}^2_m+\hat{C}^2_m+\hat{D}^2_m$. Upon taking the product of quaternionic operators $\hat{q}_1, \ldots, \hat{q}_n$, the left hand side of the inequality \eqref{eq:qF} will be a sum of four squares of average values of some observables. Then the inequality \eqref{eq:qF} is a multipartite Bell-type inequality with four observables on each site. In the case of $n=2$ it is the inequality \eqref{eq:b4}.

\begin{table}[t]
\begin{tabular}{|c|c|c|c|c|c|c|c|} \hline
          & $i_1$  & $i_2$  & $i_3$  & $i_4$  & $i_5$  & $i_6$  & $i_7$  \\ \hline
    $i_1$ & $-1$   & $i_4$  & $i_7$  & $-i_2$ & $i_6$  & $-i_5$ & $-i_3$ \\ \hline
    $i_2$ & $-i_4$ & $-1$   & $i_5$  & $i_1$  & $-i_3$ & $i_7$  & $-i_6$ \\ \hline
    $i_3$ & $-i_7$ & $-i_5$ & $-1$   & $i_6$  & $i_2$  & $-i_4$ & $i_1$  \\ \hline
    $i_4$ & $i_2$  & $-i_1$ & $-i_6$ & $-1$   & $i_7$  & $i_3$  & $-i_5$ \\ \hline
    $i_5$ & $-i_6$ & $i_3$  & $-i_2$ & $-i_7$ & $-1$   & $i_1$  & $i_4$  \\ \hline
    $i_6$ & $i_5$  & $-i_7$ & $i_4$  & $-i_3$ & $-i_1$ & $-1$   & $i_2$  \\ \hline
    $i_7$ & $i_3$  & $i_6$  & $-i_1$ & $i_5$  & $-i_4$ & $-i_2$ & $-1$   \\ \hline
\end{tabular}
\caption{The multiplication table of imaginary units of octonions.}\label{tbl:omr}
\end{table}

The inequality corresponding to the identity \eqref{eq:sq8} is
\begin{widetext}
\begin{equation}\label{eq:b8}
\begin{split}
    &\langle\hat{A}_1\hat{A}_2-\hat{B}_1\hat{B}_2-\hat{C}_1\hat{C}_2-\hat{D}_1\hat{D}_2-\hat{E}_1\hat{E}_2-\hat{F}_1\hat{F}_2-\hat{G}_1\hat{G}_2-\hat{H}_1\hat{H}_2\rangle^2 + \\
    &\langle\hat{B}_1\hat{A}_2+\hat{A}_1\hat{B}_2+\hat{D}_1\hat{C}_2-\hat{C}_1\hat{D}_2+\hat{F}_1\hat{E}_2-\hat{E}_1\hat{F}_2-\hat{H}_1\hat{G}_2+\hat{G}_1\hat{H}_2\rangle^2 + \\
    &\langle\hat{C}_1\hat{A}_2-\hat{D}_1\hat{B}_2+\hat{A}_1\hat{C}_2+\hat{B}_1\hat{D}_2+\hat{G}_1\hat{E}_2+\hat{H}_1\hat{F}_2-\hat{E}_1\hat{G}_2-\hat{F}_1\hat{H}_2\rangle^2 + \\
    &\langle\hat{D}_1\hat{A}_2+\hat{C}_1\hat{B}_2-\hat{B}_1\hat{C}_2+\hat{A}_1\hat{D}_2+\hat{H}_1\hat{E}_2-\hat{G}_1\hat{F}_2+\hat{F}_1\hat{G}_2-\hat{E}_1\hat{H}_2\rangle^2 + \\
    &\langle\hat{E}_1\hat{A}_2-\hat{F}_1\hat{B}_2-\hat{G}_1\hat{C}_2-\hat{H}_1\hat{D}_2+\hat{A}_1\hat{E}_2+\hat{B}_1\hat{F}_2+\hat{C}_1\hat{G}_2+\hat{D}_1\hat{H}_2\rangle^2 + \\
    &\langle\hat{F}_1\hat{A}_2+\hat{E}_1\hat{B}_2-\hat{H}_1\hat{C}_2+\hat{G}_1\hat{D}_2-\hat{B}_1\hat{E}_2+\hat{A}_1\hat{F}_2-\hat{D}_1\hat{G}_2+\hat{C}_1\hat{H}_2\rangle^2 + \\
    &\langle\hat{G}_1\hat{A}_2+\hat{H}_1\hat{B}_2+\hat{E}_1\hat{C}_2-\hat{F}_1\hat{D}_2-\hat{C}_1\hat{E}_2+\hat{D}_1\hat{F}_2+\hat{A}_1\hat{G}_2-\hat{B}_1\hat{H}_2\rangle^2 + \\
    &\langle\hat{H}_1\hat{A}_2-\hat{G}_1\hat{B}_2+\hat{F}_1\hat{C}_2+\hat{E}_1\hat{D}_2-\hat{D}_1\hat{E}_2-\hat{C}_1\hat{F}_2+\hat{B}_1\hat{G}_2+\hat{A}_1\hat{H}_2\rangle^2 \leqslant \\
    \langle(\hat{A}^2_1+\hat{B}^2_1&+\hat{C}^2_1+\hat{D}^2_1+\hat{E}^2_1+\hat{F}^2_1+\hat{G}^2_1+\hat{H}^2_1)(\hat{A}^2_2+\hat{B}^2_2+\hat{C}^2_2+\hat{D}^2_2+\hat{E}^2_2+\hat{F}^2_2+\hat{G}^2_2+\hat{H}^2_2)\rangle.
\end{split}
\end{equation}
\end{widetext}
It is a bipartite Bell-type inequality with eight observables on each site. To get a general multipartite inequality we need to use the algebra of octonions. It is an $8$-dimensional algebra over the reals, so each octonion $o$ can be written as
\begin{equation}\label{eq:o}
    o = x_0 + x_1 i_1 + x_2 i_2 + x_3 i_3 + x_4 i_4 + x_5 i_5 + x_6 i_6 + x_7 i_7
\end{equation}
in an unique way, where $x_l$, $l = 0, \ldots, 7$ are reals and $i_l$, $l = 1, \ldots, 7$ are imaginary units, whose multiplication rules are given by Table~\ref{tbl:omr}. The conjugation $o^*$ of the octonion \eqref{eq:o} is defined as 
\begin{equation}
    o^* = x_0 - x_1 i_1 - x_2 i_2 - x_3 i_3 - x_4 i_4 - x_5 i_5 - x_6 i_6 - x_7 i_7.
\end{equation}
The norm $|o|$ is also defined in the standard way as
\begin{equation}\label{eq:oo}
    |o| = \sqrt{o^* o} = \sqrt{x^2_0 + x^2_1 + x^2_2 + x^2_3 + x^2_4 + x^2_5 + x^2_6 + x^2_7}.
\end{equation}
The identity \eqref{eq:sq8} expresses the fact that this norm is multiplicative: $|o^{\prime} o^{\prime\prime}| = |o^{\prime}| |o^{\prime\prime}|$. This norm also satisfies the triangle inequality $|o_1 + o_2| \leqslant |o_1| + |o_2|$. We can define octonionic operators as
\begin{equation}
    \hat{o} = \hat{A} + i_1 \hat{B} + i_2 \hat{C} + i_3 \hat{D} + i_4 \hat{E} + i_5 \hat{F} + i_6 \hat{G} + i_7 \hat{H},
\end{equation}
where $\hat{A}, \ldots, \hat{H}$ are Hermitian operators. Here we identify the first imaginary unit $i_1$ with the standard complex unit $i$. For the product of the other imaginary unit with ordinary operators $\hat{f}$ we have the relation: $i_l \hat{f} = \hat{f}^\dagger i_l$, $l = 2, \ldots, 7$. But now the product is not associative, so whenever we deal with a product of more than two terms we have to explicitly group the terms.

Now we can generalize the inequality \eqref{eq:qF} as follows: if $\hat{o}_m$ are octonionic operators and $\hat{F}_m$ are Hermitian operators acting on different degrees of freedom such that $|\langle\hat{o}_m\rangle|^2 \leqslant \langle\hat{F}_m\rangle$, $m = 1, \ldots, n$, then each completely separable state satisfies the inequality
\begin{equation}\label{eq:oF}
    |\langle\hat{o}_1 \ldots \hat{o}_n\rangle|^2 \leqslant \langle\hat{F}_1 \ldots \hat{F}_n\rangle,
\end{equation}
for all $C_n$ possible groupings of the terms on the left hand side, where $C_n$ is the $n$th Catalan number defined as $C_n = (2n-2)!/(n!(n-1)!)$ (we did not show the brackets explicitly). Here we can take
$\hat{F}_m = \hat{A}^2_m+\hat{B}^2_m+\hat{C}^2_m+\hat{D}^2_m + \hat{E}^2_m + \hat{F}^2_m + \hat{G}^2_m + \hat{H}^2_m$. Upon taking the product of $\hat{o}_1, \ldots, \hat{o}_n$, the left hand side of the inequality \eqref{eq:oF} will be a sum of eight squares of average values of some observables. Then the inequality \eqref{eq:oF} is the multipartite Bell-type inequality with eight observables on each site. In the case of $n=2$ it reduces to the inequality \eqref{eq:b8}.

It is noteworthy that the inequalities \eqref{eq:b2}, \eqref{eq:b4} and \eqref{eq:b8} form a hierarchy --- the inequality \eqref{eq:b2} is a special case of \eqref{eq:b4}, which is in turn a special case of \eqref{eq:b8}. As shown in \cite{prl-99-210405}, the inequality \eqref{eq:b2} can be violated. Thus it is clear that our inequalities \eqref{eq:b4} and \eqref{eq:b8} can be violated as well. Any violation of these inequalities is a clear signature of entanglement.

The inequalities \eqref{eq:b2}, \eqref{eq:b4} and \eqref{eq:b8} can be also obtained in another way. The integral form of the inequality \eqref{eq:px} reads as
\begin{equation}\label{eq:X}
    \left|\int p(\lambda)X(\lambda)d\lambda\right|^2 \leqslant \int p(\lambda)|X(\lambda)|^2d\lambda,
\end{equation}
where $p(\lambda)$ is a probability distribution on a measurable set $\Lambda$, and $X(\lambda)$ is a real- or complex-, quaternion- or octonion-valued function of $\lambda \in \Lambda$. The set $\Lambda$ can be thought of as a set of hidden variables, which completely specify the state under study. The inequality \eqref{eq:X} simply states that $|\langle X \rangle|^2 \leqslant \langle|X|^2\rangle$. Let us take instead of $X$ the operator $\hat{X}$, which is a product of ordinary non-Hermitian operators $\hat{f}_m$, quaternionic operators $\hat{q}_m$ or octonionic operators $\hat{o}_m$ defined in \eqref{eq:oo}, acting on different degrees of freedom $m = 1, \ldots, n$, and find the quantum mechanical analogue of the quantity $|X|^2$. For example, in the case of ordinary non-Hermitian operators, we have $\hat{f}^\dagger_m\hat{f}_m = \hat{A}^2_m+\hat{B}^2_m+i[\hat{A}_m, \hat{B}_m]$. In a local hidden variable theory all commutators must be zero, so $|X|^2$ must be replaced by the product $\prod^n_{m=1}(\hat{A}^2_m + \hat{B}^2_m)$. Analogously, in the case of quaternionic operators $|X|^2$ must be replaced by $\prod^n_{m=1}(\hat{A}^2_m + \hat{B}^2_m + \hat{C}^2_m + \hat{D}^2_m)$ and in the case of octonionic case by $\prod^n_{m=1}(\hat{A}^2_m + \ldots + \hat{H}^2_m)$. Then we again arrive to the inequalities \eqref{eq:b2}, \eqref{eq:b4} and \eqref{eq:b8} respectively. This approach is simpler then the one we started with, but it is ambiguous and does not give strictly relate the obtained inequalities to separability.

In conclusion, we have obtained Bell-type inequalities for observables with a general spectrum. They apply to measurements of up to eight observables, for arbitrary systems.
The multipartite forms of the inequalities are related to the algebras of quaternions and octonions.

The authors gratefully acknowledge support by the Deutsche Forschungsgemeinschaft.

\end{document}